# Hybrid Mortality Prediction using Multiple Source Systems

Isaac Mativo[1], Yelena Yesha[1], Michael Grasso[2], Tim Oates[1], Qian Zhu[3]

[1]Department of Computer Science, University of Maryland Baltimore County (UMBC)

[2]School of Medicine, University of Maryland Baltimore (UMB)

[3]Department of Information Systems, University of Maryland Baltimore County (UMBC)

*Abstract* — The use of artificial intelligence in clinical care to improve decision support systems is increasing. This is not surprising since by its very nature, the practice of medicine consists of making decisions based on observations from different systems both inside and outside the human body. In this paper, we combine three general systems (ICU, diabetes, and comorbidities) and use them to make patient clinical predictions. We use an artificial intelligence approach to show that we can improve mortality prediction of hospitalized diabetic patients. We do this by utilizing a machine learning approach to select clinical input features that are more likely to predict mortality. We then use these features to create a hybrid mortality prediction model and compare our results to non artificial intelligence models. For simplicity, we limit our input features to patient comorbidities and features derived from a well-known mortality measure, the Sequential Organ Failure Assessment (SOFA).

I. Introduction

Clinical decision support systems are driven by large amounts of data sourced from various disparate systems. These may include data from lab instruments, wearable devices, prior hospital records, patient reported data, etc. Often, the clinician is exposed to a lot of this data and is expected to make decisions to improve clinical outcomes. Although decision support systems are effective in logical tasks such as alerting the user when some parameters are outside defined ranges, they are not designed to learn from the data they have. As such, they are limited to what they have been programmed to do. This limitation becomes an opportunity to experiment with artificial intelligence within the clinical setting [1]. Specifically, a hybrid system that combines attributes from both traditional clinical decision support systems and attributes derived from machine learning tools to create a prediction model becomes attractive. Hybrid systems have been used in many fields to improve performance in areas such as managing humanitarian relief chains [2] and multiple-criteria decision-making [3]. In healthcare, hybrid decision support systems have been used to help clinicians make more informed decisions [4][5]. For mortality prediction, several hybrid approaches have been proposed such as the use of hybrid image features on neural network for cancer patients [3][6]. In this paper, we take a systems theory view of a clinical setting whereby many different systems come together to inform the clinician on the best decisions to make. In the next section, we will describe what we consider to be the three systems that we consider and apply machine learning tools on.

II. The Three Systems

### A. Diabetes

For this paper, we consider the chronic disease diabetes as one of our three systems. This is because diabetes, being a disorder of metabolism, affects multiple systems within the human body [7]. When blood sugar is not well controlled and therefore outside the normal range for prolonged periods of time, the proper functioning of various organ systems is compromised. Consequently, this manifests in various body conditions such as kidney disease, neuropathy, liver disease, etc. [8].

From a cybernetics perspective, diabetes is interesting because it involves the delicate control of blood sugar by both systems inside and outside of the body. Diabetic patients often rely on external interventions such as medication ingestion, insulin injection, dietary controls, and lifestyle changes such as exercise. When the sugar level in blood is found to be outside the normal range based on information gathered, various actions are taken to bring it under control.

### B. Intensive Care Unit

We consider the Intensive Care Unit (ICU) in a hospital to be our next system. This is because the ICU is a specialized facility with many controlled and interconnected systems that work together to intervene in critically ill patients. In the ICU, information is availed through various channels such as electronic medical records, clinical notes, instrument measurements, etc. [8] [9] All this information is then used to provide a level of clarity in making clinical decisions. Various aspects of this information such as temporal occurrence (e.g. did the blood pressure fall after or before takings a specific medication?) dependency (e.g. eat solid foods before taking a certain medication), are important. As such, in an ICU facility, the proper control of information from various systems is critical [10]. In this paper therefore, we view the ICU facility as one large system with many subsystems within it.

### C. Comorbidities

Comorbidities are co-occurring disease conditions within a person. Often, diabetic patients have other disease conditions as well. Some of these conditions are more commonly associated with diabetes than others [11]. In this paper, we think of comorbidities as a system because they are well-defined conditions within the human body that individually and collectively influence the wellbeing of the entire body. Since comorbidities are an important piece of information considered in patient treatment, we considered comorbidities to be relevant in informing our hybrid prediction model.

With these three systems, we build a hybrid mortality prediction model and evaluated its performance. It is a hybrid model because it combined both machine learning techniques and non-machine learning techniques to select it features and perform its predictions. In the next section, we discuss the materials we used to do our experiment.

III. Materials

A. *Medical Information Mart for Intensive Care (MIMIC) III*

The data source we used to get our patient population is MIMIC-III [12]. MIMIC-III is a large and single-center database comprising de identified health related information containing 46,520 patients of whom 20,399 are female and 26,121 are male. These patients stayed in critical care units of the Beth Israel Deaconess Medical Center between 2001 and 2012. The database includes information such as demographics, vital sign measurements, laboratory test results, procedures, medications, caregiver notes, imaging reports, and mortality. In this study, we accessed MIMIC-III to extract diabetes patients for mortality prediction.

B. Elixhauser *comorbidity measure.*

In order to get the comorbidity information for our diabetic patients, we used The Elixhauser comorbidity index[13]. This index consists of 30 comorbidity measures that have been shown to have an impact on patient hospital length of stay, cost, and mortality [14]. Classification into this index is based on conditions described by the ICD-9-CM (International Classification of Diseases, Ninth Edition, Clinical Modifications) discharge records. The Elixhauser index assigns a value of 1 for each of 30 predefined conditions, and computes the sum. Therefore, a score of 10 indicates a patient has 10 conditions present.

### C. Mortality Measure

To compare our hybrid model prediction accuracy with a needed a tool that measured mortality. For this, we chose the Sequential Organ Failure Assessment (SOFA) [15]. SOFA is developed from a large sample of ICU patients from around the world and gives a score based on the severity of on a patient. The SOFA score is computed based on six variables, with each variable corresponding to an organ system. The organ systems included in SOFA are respiratory, cardiovascular, hepatic, coagulation, renal and neurological systems. Each organ system is assigned a value between 0 (good) and 4 (severe). The SOFA score is the sum of all the individual organ system values.

IV. Methods

In this section, we describe the methods we used to retrieve the patient cohort, computer the mortality and comorbidity measure, and to predict mortality. For this study, ethical approval was not needed because the data used is de-identified to conform with the Health Insurance Portability and Accountability Act (HIPAA). Additionally, the data is publicly available to researchers.

### A. Patient Retrieval

From our MIMIC III PostgreSQL database, we retrieved the diabetes patient cohort from MIMIC III database based on ICD-9 codes for diabetes. To avoid restricting diabetic patients with comorbidities for further mortality prediction, we extracted the entire diabetes diagnoses for each of the diabetic patients.

### B. Mortality Measure Computation

We used the SOFA score as a mortality measure. To calculate the SOFA score for each patient, we extracted the patient in the PostgreSQL database for calculating the the SOFA score corresponding to the organ areas as shown in table 1. We extracted this data using query scripts in PostgreSQL from the MIMIC III diabetic patients data based on the first day of each patient's' ICU stay. For each patient, we used an SQL script to calculate the actual points for each organ system, then added up the points for each organ system to obtain the SOFA score. For missing values, we assigned zero points to that organ system. We did this for all collected diabetic patients.

TABLE 1: SOFA ORGAN AREAS AND POINTS THRESHOLDS

| Organ/Points | 1 | 2 | 3 | 4 |
|---|---|---|---|---|
| **Respiratory** $PaO_2/FiO_2$ (mmHg) | <400 | <300 | <200 and respiratory support | <100 and respiratory support |
| **Cardiological** Mean arterial pressure | MAP < 70 mm/Hg | dop <= 5 or dob | dop > 5 OR epi <= 0.1 OR nor <= 0.1 | dop > 15 OR epi > 0.1 OR nor > 0.1 |
| **Renal** Creatinine (mg/dl) [μmol/L] (or urine output) | 1.2–1.9 [110-170] | 2.0–3.4 [171-299] | 3.5–4.9 [300-440] (or < 500 ml/d) | > 5.0 [> 440] (or < 200 ml/d) |
| **Hepatic** Bilirubin (mg/dl) [μmol/L] | 1.2–1.9 [> 20-32] | 2.0–5.9 [33-101] | 6.0–11.9 [102-204] | > 12.0 [> 204] |
| **Neurological** Glasgow coma scale | 13 - 14 | 10 - 12 | 6 - 9 | <6 |
| **Coagulation** Platelets×103/μl | <150 | <100 | <50 | <20 |

C. Comorbidity Measure and Weight Computation

We used the Elixhauser index to calculate the comorbidity measure for each diabetic patient in MIMIC-III. For each diabetic patient, we extracted 29 different diagnosis excluding diabetes and quantified them by using the Elixhauser index for comorbidities represented in the diabetic population.

In order to improve the efficiency of our classifier by reducing the number of features, we identified the top five comorbidities that are most predictive of mortality across the identified diabetic patients. To do this, we experimented using 5 different feature selection algorithms, which are Gain Ratio [16], Correlation [17], Symmetrical Uncertainty [18], Information Gain [19], and Correlation Feature Selection (CFS) Subset Evaluator [20]. For each algorithm run, we ranked the comorbidities based on the resulted predictive mortality. We then averaged these results over the 5 algorithms to get our final top 5 most predictive comorbidities. The final list of the comorbidities was cardiac arrhythmias, coagulopathy, metastatic cancer, congestive heart failure, and fluid electrolyte.

### D. Mortality Prediction

For mortality prediction, we created a hybrid mortality prediction model that took as input the 6 SOFA scores representing the 6 organ systems and the top 6 ranked comorbidities, for a total of 11 features. We then experimented with a Naive Bayes classifier and a Random Forest classifier with 10-fold cross validation across our patient population. We compared the prediction accuracy of our hybrid system with the accuracy of (i) a prediction model with only the SOFA scores and (ii) a prediction model with only the comorbidity information. Our accuracy comparison was based on the AUC on both the Naive Bayes and Random Forest classifiers on mortality.

## V. Results

### A. Patient Retrieval

A total of 10,403 of diabetes patients were retrieved from 46,520 patients contained in MIMIC III database. Of all the identified diabetic patients, 1,513 died while in ICU. Among the 10,403 diabetes patients, we identified 29 different comorbidities.

B. Mortality Measure Computation

Table 2 shows the calculated SOFA scores and the corresponding mortality rates as computed from our data. Our mortality rates were consistent with the expected SOFA mortality rates.

TABLE 2: CALCULATED MORTALITY RATES

| SOFA Score | Calculated Mortality Rate |
|---|---|
| 0 to 6 | 7% |
| 7 to 9 | 20% |
| 10 to 12 | 39% |
| 13 to 14 | 59% |
| 15 | 64% |
| 16 to 24 | 72% |

C. Comorbidity Measure and Weights

Figure 1 shows the results of calculated comorbidity for the diabetic patients using the Elixhauser index. As can be seen, most of the patients had comorbidities.

FIGURE 1: COMORBIDITIES ACROSS PATIENTS

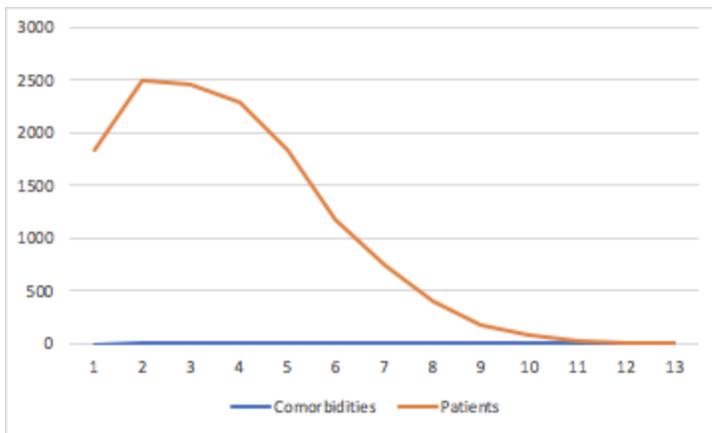

Table 3 shows results generated via the five feature extraction algorithms to determine what comorbidities to be applied for prediction classifiers. These are the comorbidities that were used as features in our model.

TABLE 3: HIGHLY WEIGHED COMORBIDITIES

| Comorbidity | Average Rank |
|---|---|
| Coagulopathy | 1.6 |
| Cardiac arrhythmias | 3 |
| Fluid electrolyte | 3.2 |
| Metastatic cancer | 3.8 |
| Congestive heart failure | 5.2 |

D. Mortality Prediction

Table 4 shows the results generated by the Naive Bayes and Random Forest classifiers. The ROC is lowest when using only the 5 comorbidity features are used, is better when using the 6 SOFA scores are used, and best when both the 5 comorbidities and 6 SOFA scores are combined. Thus, the hybrid model outperforms the natively SOFA or comorbidity only models.

TABLE 4: MORTALITY PREDICTION RESULTS

| Features (count) | Random Forest ROC Area | Naive Bayes ROC Area |
|---|---|---|
| Comorbidities (5) | 0.667 | 0.672 |
| SOFA (6) | 0.731 | 0.743 |
| Combined (11) | 0.763 | 0.772 |

VI. Discussion

In this study, we proposed viewing ICU clinical care through the lens of cybernetics. We identified three systems; the ICU, diabetes, and patient comorbidity, as related and defined systems that are important in supporting clinical decisions. We examined the mortality impact of comorbidities on diabetic patients patients admitted in an ICU unit using machine learning tools. We created a hybrid mortality prediction model and compare its results to non-hybrid models. The results we obtained support our claim that mortality prediction can be improved by incorporation key features from different systems (comorbidities in our case) and using artificial intelligence to train our classifiers.

Another key observation from our results is that we can still get improved performance from our prediction model without using all of the available input features. In our case, we used statistical techniques to rank all our comorbidities based on their weight in impacting mortality. This observation is important because within systems, we can optimize performance by carefully selecting the components that affect our output. This careful selection results in dimension reduction and consequently reduces computational resources needed to run algorithms while at the same time producing improved results.

A logical extension of this research is further defining the subsystems within our system, or yet still identifying other systems that are important in improving clinical decision support systems. An improvement of our approach would be to take a closer look at the information flow within and across our systems and evaluate its impact on the overall efficiency of an ICU setting. Important makers, in this case, could include items such as readmission rates, adverse events, hospital length of stay, central line infection, etc. We hope that with this study we have demonstrated that it is possible to view clinical decision support systems from cybernetics perspective and encouraged other researchers to consider this approach.

DISCLOSURE OF INTERESTS

The authors report no conflict of interest.